\documentclass[smallextended]{svjour3}       % onecolumn (second format)
\usepackage[sort,numbers]{natbib}
\smartqed  % flush right qed marks, e.g. at end of proof
\usepackage{graphicx}
\usepackage{todonotes}
\usepackage{framed}
\usepackage[hidelinks]{hyperref}

\usepackage{etoolbox}

\makeatletter
\def\cl@chapter{\@elt {theorem}}
\makeatother
\usepackage{cleveref}

\def\addvalue#1#2{\expandafter\gdef\csname my@data@#1\endcsname{#2}}
\def\usevalue#1{\csname my@data@#1\endcsname}
\newcommand{\mylabel}[2]% #1 = label name, #2 = text for \ref
{\label{#1}\addvalue{#1}{#2}}
\newcommand{\mynameref}[1]{\usevalue{#1}}

\newcounter{RQCounter}

\crefname{RQCounter}{RQ}{RQs}

\newcommand{\newrquestion}[2]{%
\refstepcounter{RQCounter}%
\mylabel{#1}{#2}%
\rquestion{#1}%
}

\newcommand{\rquestion}[1]{
  \noindent\textbf{RQ{\ref{#1}}}: \rqn{#1}
}

\newcommand{\rquestionbox}[1]{%
\vspace{2ex}%
\noindent\fbox{\parbox{\linewidth}{{\emph{\rquestion{#1}}}}}%
\vspace{2ex}%
}

\newcommand{\rqn}[1]{\mynameref{#1}}

\newcommand{\Website}[1]{Website visited on September 16, 2020: {\url{#1}}.}
\newcommand{\website}[1]{\Website{#1}}

\usepackage{ifthen}
\usepackage{xspace}
\newboolean{anonymous}
\setboolean{anonymous}{false} %% <-- Set boolean here!
\newcommand{\ING}{ING\xspace}
\ifthenelse{\boolean{anonymous}}{
  \renewcommand{\ING}{ABC\xspace}
}{}

\newcommand{\highlight}[1]{\begin{framed}%
  \noindent\emph{#1}
\end{framed}}

\renewcommand{\today}{March 1, 2021}

\journalname{Empirical Software Engineering}

\begin{document}

% \title{An Exploratory Study on the Lifecycle of Machine Learning Applications in Fintech}
% \title{The Life of Machine Learning Applications in Fintech}
% \title{The Untold Story of Machine Learning Applications in Fintech: beyond fancy algorithms}
% \title{Machine Learning in Fintech: beyond fancy algorithms}
% \title{The Helicopter View of Machine Learning Applications in Fintech}
% \title{A bird's eye on Machine Learning Applications in Fintech}
% \title{Machine Learning Beyond Fancy Algorithms: An Exploratory Study in Fintech}
% \title{Machine Learning Behind The Scenes:\\ An Exploratory Study in Fintech}
\title{AI Lifecycle Models Need To Be Revised}
\subtitle{An Exploratory Study in Fintech}

\author{Mark Haakman        \and
        Lu\'{i}s Cruz       \and
        Hennie Huijgens     \and
        Arie van Deursen    \and
}


\institute{Mark Haakman \at
              AI For Fintech Research, ING \\
              \email{Mark.Haakman@ing.com}           %  \\
           \and
           Lu\'{i}s Cruz \at
              Delft University of Technology\\
              \email{l.cruz@tudelft.nl}
              \and
           Hennie Huijgens \at   
              AI For Fintech Research, ING \\
              \email{Hennie.Huijgens@ing.com}           %  \\
           \and
           Arie van Deursen \at
              Delft University of Technology\\
              \email{Arie.vanDeursen@tudelft.nl}
}

\date{Received: date / Accepted: date}

\maketitle

%!TEX root = ../main.tex

\begin{abstract}
% Background
% Artificial Intelligence has become increasingly important for organizations.

Tech-leading organizations are embracing the forthcoming artificial
intelligence revolution. Intelligent systems are replacing and cooperating with
traditional software components. Thus, the same development processes and
standards in software engineering ought to be complied in artificial
intelligence systems.
% Goals
This study aims to understand the processes by which artificial
intelligence-based systems are developed and how state-of-the-art lifecycle
models fit the current needs of the industry.
% Methods
We conducted an exploratory case study at \ING, a global bank with a strong
European base. We interviewed 17 people with different roles and from different
departments within the organization.
% Results
We have found that the following stages have been overlooked by previous
lifecycle models: \emph{data collection}, \emph{feasibility study},
\emph{documentation}, \emph{model monitoring}, and \emph{model risk assessment}.
% Conclusions
Our work shows that the real challenges of applying Machine Learning go much
beyond sophisticated learning algorithms -- more focus is needed on the entire
lifecycle. In particular, regardless of the existing development tools for
Machine Learning, we observe that they are still not meeting the
particularities of this field.

%The key challenges of developing Machine Learning applications in
%fintech organizations are model governance and technology access.

\end{abstract}

%!TEX root = ../main.tex

\section{Introduction} \label{sec:introduction}
% Importance of AI and bridge SE to AI and AI to ML.
Artificial Intelligence~(AI) has become increasingly important for organizations to support
customer value creation, productivity improvement, and insight discovery. Pioneers in the AI
industry are asking how to better develop and maintain AI software~\citep{menzies2019five}. This
paper focuses on Machine Learning, the branch of AI that deals with the automatic generation of
knowledge models based on sample data. Throughout
this article, Machine Learning and AI are used interchangeably.

% Background
Although most of the AI techniques are not so recent (e.g., neural networks
were already being applied in the 1980s~\citep{mead1989analog}), the recent
access to large amounts of data and more computing power has exploded the
number of scenarios where AI can be
applied~\citep{wu2019machine,bernardi2019150}. In fact, AI is now being used to
add value in critical business scenarios. Consequently, a number of new
challenges are emerging in the lifecycle of AI systems, comprising all the
stages from their conception to their retirement and disposal. Like normal
software applications, these projects need to be planned, tested, debugged,
deployed, maintained, and integrated into complex systems.

Companies leading the advent of AI are reinventing their development processes and coming up with
new solutions. Thus, there are many lessons to be learned to help other organizations and guide
research in a direction that is meaningful to the industry. This is particularly relevant for
heavily-regulated industries such as fintech. Industries in the fintech domain ought to make sure
that not only they adhere to ever-changing regulations\footnote{\emph{Bank regulations change every 12 minutes} by Chris M. Skinner (2017). Retrieved on \today: \url{https://thefinanser.com/2017/01/bank-regulations-change-every-12-minutes.html/}} but also that the usage of their products is compliant. Hence, new processes need to be designed to make sure AI systems meet all required
standards.

Recent research has addressed how developing AI systems is different from
developing regular Software Engineering systems. A case study at Microsoft
identified the following differences~\citep{amershi2019software}: 1) data
discovery, management, and versioning are more complex; 2) practitioners ought
to have a broader set of skills; and 3) modular design is not trivial since AI
components can be entangled in complex ways. Unfortunately, existing research
offers little insight into the challenges of transforming an existing IT
organization into an AI-intensive one.

Examples of existing models describing the Machine Learning lifecycle are the
Cross-Industry Standard Process for Data Mining
(CRISP-DM)~\citep{shearer2000crisp} and the Team Data Science Process
(TDSP)~\citep{TDSP}. However, Machine Learning is being used for different
problems across many different domains. Given the fast pace of change in AI and
recent advancements in Software Engineering, we suspect that there are
deficiencies in these lifecycle models when applied to a fintech context.

To remedy this, we set out this exploratory case study aimed at understanding
and improving how the fintech industry is currently dealing with the challenges
of developing Machine Learning applications at scale. \ING{} is a relevant case
to study, since it has a strong focus on financial technology and Software
Engineering and it is undergoing a bold digital transformation to embrace
AI as an important competitive factor. By studying \ING{},
we provide a snapshot of the rapid evolution of the approach to Machine Learning development.

We define the following research questions for our study:

\vspace{1ex}
\noindent\emph{\newrquestion{rq:lifecycle}{How do existing Machine Learning lifecycle models fit the fintech domain?}}

\vspace{1ex}
\noindent\emph{\newrquestion{rq:challenges}{What are the specific challenges of
developing Machine Learning applications in fintech organizations?}}
\vspace{1ex}

We interviewed 17 people at \ING with different roles and from different departments.
Thereafter, we triangulated the resulting data with other resources available
inside the organization. Furthermore, we refine the existing lifecycle models
CRISP-DM and TDSP based on our observations at \ING.

Our results unveil important challenges that ought to be addressed when
implementing Machine Learning at scale. Feasibility assessments, documentation,
model risk assessment, and model monitoring are stages that have been overlooked by
existing lifecycle models. There is a lack of standards and there is a need for
automation in the documentation and governance of Machine Learning models.
Finally, we pave the way for shaping the education of AI to address the current
needs of the industry.

The remainder of this paper is structured as follows. In
Section~\ref{sec:background} we introduce existing lifecycle models and
describe related work. In Section \ref{sec:design}, we outline the study
design. We report the data collected in Section~\ref{sec:datanalysis} and we
answer the research questions in Section~\ref{sec:data_synthesis}. We discuss
our findings and threats to validity in
Section~\ref{sec:discussion}. Finally, in section~\ref{sec:conclusions}, we
pinpoint conclusions and outline future work.

%!TEX root = ../main.tex

\section{Background}\label{sec:background}

In this section, we present the lifecycle models considered in this study and
examine existing literature outlining the differences with our study.

\subsection{Existing Lifecycle Models} \label{sec:existing}
In this study, we consider three reference models for the lifecycle of Machine
Learning applications: Cross-Industry Standard Process for Data Mining
(CRISP-DM)~\citep{shearer2000crisp}, the Team Data Science Process
(TDSP)~\citep{TDSP}, and the Microsoft model described by \citet{amershi2019software}. We chose CRISP-DM, as although it is twenty years old, it is still the \textit{de facto}
standard for developing data mining and knowledge discovery
projects~\citep{martinez2019crisp}. We selected TDSP as modern industry
methodology, which has at a high level much in common with CRISP-DM. Finally, we also include the model described by \citeauthor{amershi2019software}, which is based on CRISP-DM and TDSP and addresses the workflow of software engineering teams~\citep{amershi2019software}. There are
other methodologies, but most are similar to these three. Findings in our
paper can be extrapolated to those other lifecycle models.

\begin{figure}
  \centering
  \includegraphics[width=0.6\textwidth]{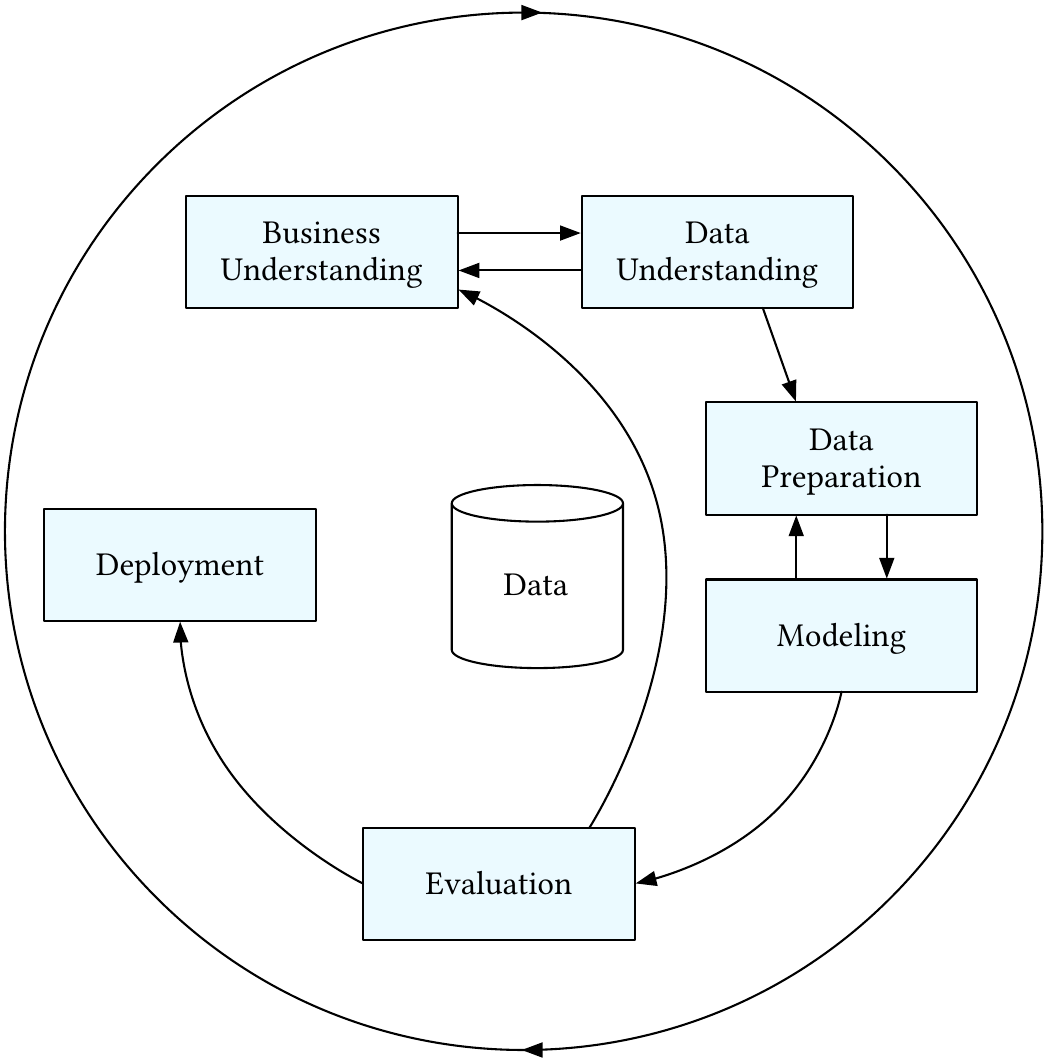}
  \caption{Cross-Industry Standard Process for Data Mining (CRISP-DM).}
  \label{fig:crisp-dm}
\end{figure}

CRISP-DM aims to provide anyone with ``a complete blueprint for conducting a
data mining project''~\citep{shearer2000crisp}. Although data mining is not the
common term used nowadays, it is valid for any project applying scientific
methods to extracting value from data, including Machine Learning~\citep{martinez2019crisp}.
CRISP-DM breaks down a project into six phases, as presented in
Fig.~\ref{fig:crisp-dm}. It typically starts with \textbf{Business
Understanding} to determine business objectives, going back and forward with
\textbf{Data Understanding}. It is followed by \textbf{Data Preparation} to
make data ready for \textbf{Modeling}. The produced model goes through an
\textbf{Evaluation} in which it is decided whether the model can go for
\textbf{Deployment} or it needs another round of improvement. The arrows
between stages indicate the most relevant and recurrent dependencies, while the
arrows in the outer circle indicate the evolution of Machine Learning systems after being
deployed and their iterative nature.

Based on CRISP-DM, a number of lifecycle models have been
proposed~\citep{martinez2019crisp,mariscal2010survey} to address varying
objectives. Derived models extend CRISP-DM for projects with geographically
dispersed teams~\citep{moyle2001ramsys}, with large amounts of data and more
focus on automation~\citep{wu2013data,rollins2015foundational}, or targeting the
model reuse across different contexts~\citep{martinez2017casp}.

\begin{figure}
  \centering
  \includegraphics[width=0.8\textwidth]{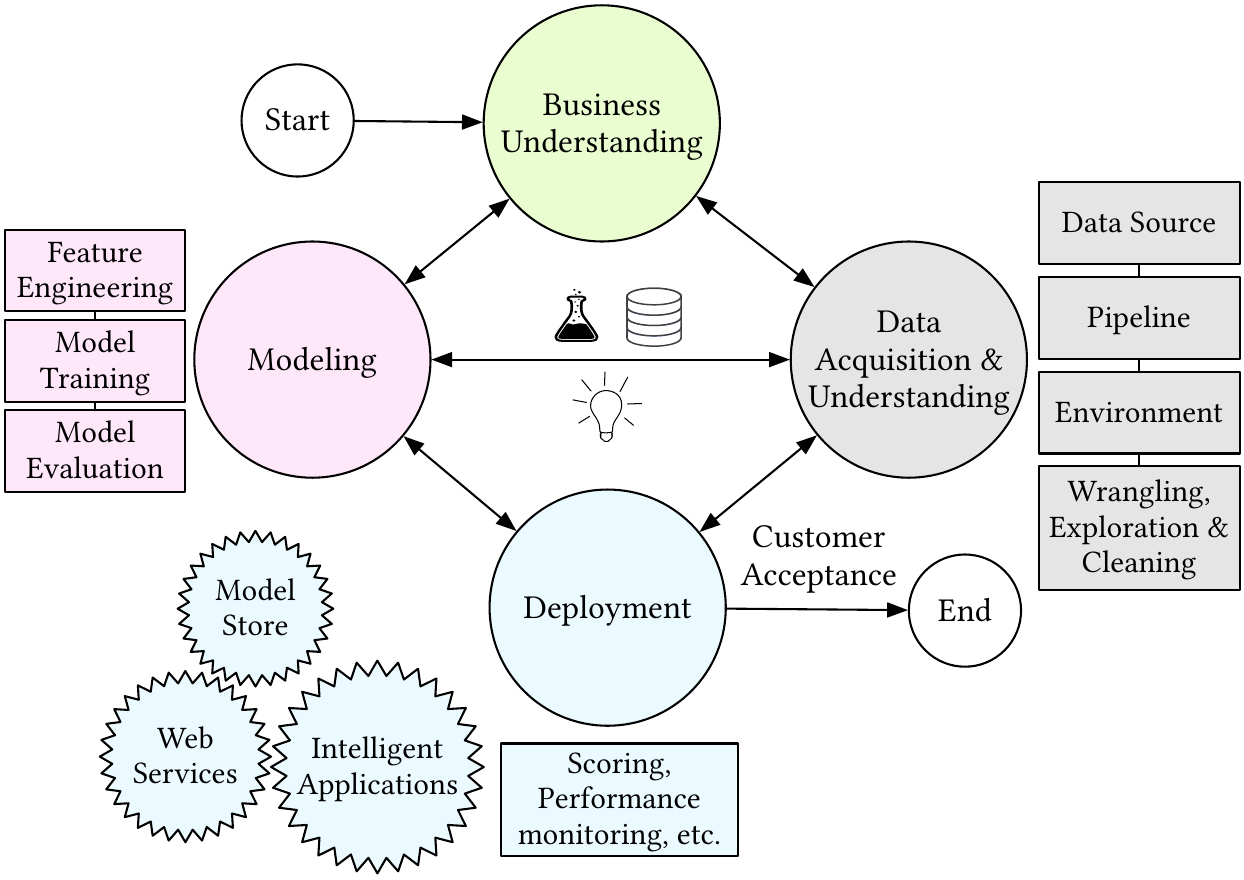}
  \caption{Team Data Science Process (TDSP).}
  \label{fig:tdsp}
\end{figure}

TDSP is ``an agile, iterative data science methodology'' proposed by Microsoft, to deliver Machine
Learning solutions efficiently~\citep{TDSP}. The original methodology includes
four major stages, as can be seen in Fig.~\ref{fig:tdsp}: \textbf{Business
Understanding}, \textbf{Data Acquisition}, \textbf{Modeling} and
\textbf{Deployment}. As depicted by the arrows in the figure, TDSP proposes
stronger dependencies but does not enforce a particular order between stages,
emphasizing that different stages can be iteratively repeated at almost any
time in the project.

\begin{figure}
  \centering
  \includegraphics[width=1.0\textwidth]{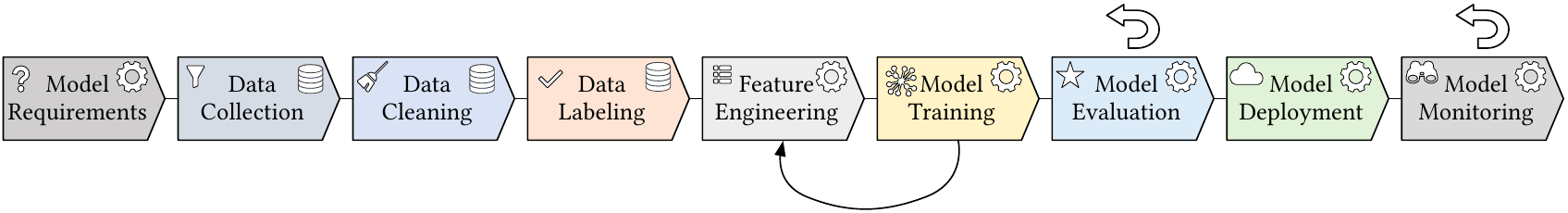}
  \caption{Microsoft's machine learning workflow described by \citet{amershi2019software}.}
  \label{fig:amershi}
\end{figure}

\citet{amershi2019software} describe the nine stages followed by software engineering teams at Microsoft who are integrating machine learning into application and platform development. The workflow is presented in Fig.~\ref{fig:amershi}, with nine stages: \textbf{Model Requirements}, \textbf{Data Collection}, \textbf{Data Cleaning}, \textbf{Data Labeling}, \textbf{Feature Engineering}, \textbf{Model Training}, \textbf{Model Evaluation}, \textbf{Model Deployment}, and \textbf{Model Monitoring}. The large feedback arrows in the figure depict stages that can be followed by any of their precedent stages. It is the case of Model Evaluation and Model Monitoring. The smaller feedback arrow shows that Model Training and Feature Engineering are typically revisited iteratively.

Despite the number of advancements proposed in previous work, we argue that they do not
tackle AI systems that target challenges faced by the fintech industry.
Our work pinpoints the changes that needed to be accommodated for AI systems operating under heavy-regulated scenarios and bringing value over pre-existing
non-data-driven approaches.

\subsection{Related Work} \label{sec:rw}

% lifecycle (case) studies.
The Machine Learning development lifecycle has been studied in practice in previous research.
\citet{amershi2019software} have conducted a case study at Microsoft to study the differences
between Software Engineering and Machine Learning. 
They interviewed 15 software engineers and a conducted a survey with 551 software engineers, yielding 4 main contributions: 1) a description of a machine learning workflow, that we use for comparison with our study; 2) a set of best practices for building applications with machine learning models, 3) a preliminary maturity model for teams developing machine learning applications, and 4) a discussion of the fundamental differences between developing software systems that integrate machine learning models and traditional software systems.

According to \citet{amershi2019software}, the main differences that set machine learning apart from traditional software
engineering can be summarized as follows: a) handling the data needed for Machine Learning
applications is considerably more complex, b) model customization and reuse require a wide set of
skills that are not typically found in software teams, c) it is hard to isolate two different
machine learning models that operate in the same system -- often they ought to be developed and
training together. We complement this study by not restricting our observations to software
engineering teams. Hence, we include any practitioner working on a machine learning system. Moreover, we argue that Microsoft as a long history in developing machine learning systems, which might neglect some of the challenges that organizations shifting to AI have to endure.
Finally, we compare our observations with existing Machine Learning lifecycle models, including
the one proposed by \citeauthor{amershi2019software}.

Another case study from industry has been performed at Booking.com by Bernardi et
al.~\citep{bernardi2019150}. In contrast with academic research in which Machine Learning models
are validated by means of an error measurement, models at Booking.com are validated through
business metrics such as conversion or cancellations. The paper describes process stages such as
model designing, deployment, monitoring, and evaluation, but no formal lifecycle model is defined.
Moreover, we hypothesize that the fintech domain poses extra challenges that stem from having to adhere to heavy regulations.

Hill et al.~\citep{hill2016trials} studied how people develop intelligent
systems in practice. The study leverages a high-level model of the process and
identifies the main challenges. Results show that developers struggle with
establishing repeatable processes and that there is a basic mismatch between
the tools available versus the practical needs. In this study, we extend the
work by Hill et al. by looking more closely at what happens after the Machine Learning model
has been evaluated, for example regarding its deployment and monitoring.

The paper by Lin and Ryaboy~\citep{lin2013scaling} describes the \textit{big
data mining cycle} at Twitter, based on the experience of the two authors. The
main points made are that for data-driven projects, most time goes to
preparatory work before, and engineering work after the actual model training
and that a significant amount of tooling and infrastructure is required. In our
study, we validate the recommendations of these two experts with a case study
with seventeen participants.

% Studies reporting challenges.
Concrete challenges data scientists face are elaborated upon in the study by
Kim et al.~\citep{kim2017data}. They have surveyed 793 professional data
scientists at Microsoft. An example of a challenge found is that the
proliferation of data science tools makes it harder to reuse work across teams.
This challenge is also reinforced in the study by Ahmed et
al.~\citep{ahmed2019machine}. As models are mostly implemented without standard
API, input format, or hyperparameter notation, data scientists spend
considerable effort on implementing glue code and wrappers around different
algorithms and data formats to employ them in their pipelines. Ahmed et
al.~\citep{ahmed2019machine} show evidence that most models need to be rewritten
by a different engineering team for deployment. The root of this challenge lies
on runtime constraints, such as a different hardware or software platform, and
constraints on the pipeline size or prediction latency.

% Bringing SE to ML.
More studies looked at Machine Learning from a Software Engineering
viewpoint. Sculley et al.~\citep{sculley2015hidden} identified a number of
Machine Learning-specific factors that increase technical debt, such as
boundary erosion and hidden feedback loops. Breck et al.~\citep{breck2017ml}
have proposed 28 specific tests for assessing production readiness for Machine
Learning applications. These tests include tests for features and data, model
development, infrastructure, and monitoring. Arpteg et
al.~\citep{arpteg2018software} have identified Software Engineering challenges
of building intelligent systems with deep learning components based on seven
projects from companies of different types and sizes. These challenges include
development, production, and organizational challenges, such as experiment
management, dependency management, and effort estimation. In this current
study, we will extend this line of research and identify where Software
Engineering can help mitigate inefficiencies in the development and evolution
of Machine Learning systems.

%!TEX root = ../main.tex

\section{Research Design} \label{sec:design}
To identify the gaps in the existing Machine Learning
lifecycle models and explore key challenges in the field, we perform a single-case exploratory case
study. This is a recurrent methodology to define new research by looking at
concrete situations and to shed empirical light on existing concepts and
principles~\citep{yin2017case}. We follow the guidelines proposed by
Brereton et al.~\citep{brereton2008using} and Yin's~\citep{yin2017case} case study
methodology.

It is not our objective to build an entirely new theory from the ground up.
For that reason, we do not
adopt a Grounded Theory (GT) approach, although we do use a number of
techniques based on GT~\citep{stol2016grounded}: e.g., theoretical sampling,
memoing, memo sorting, and saturation.

The design of the study is further described in this section.

\subsection{The Case}
% Describe the company
\ifthenelse{\boolean{anonymous}}{ % Anonymous paragraph.
The case under study is \ING, a global bank with a strong European base. \ING
offers retail and wholesale banking services to XX million customers in over
XX countries, with over XX,000 employees. \ING has a strong focus on fintech, the digital transformation of the financial sector, and
professionalization of AI development.
}{ % Paragraph mentioning ING and real numbers.
The case under study is \ING, a global bank with a strong European base. \ING
offers retail and wholesale banking services to 38 million customers in over
40 countries, with over 53,000 employees~\citep{ING_stats}. \ING has a strong
focus on fintech, the digital transformation of the financial sector, and
professionalization of AI development.
}

% Describe use-cases for Machine Learning
A bank of this size has many use cases where Machine Learning can help. Examples include
traditional banking activities such as assessing credit risk, the execution of
customer due diligence and transaction monitoring requirements related to
fighting financial economic crime. Other examples of use cases are improving
customer service and IT infrastructure monitoring.

% Describe the way of working
% Development teams at \ING follow an agile way of working and the organization is structured
% similar to the Spotify organization model~\citep{kniberg2012scaling} with
% tribes, squads, and chapters. The basic unit is a \textit{squad}, a
% self-organizing team similar to a Scrum team. A collection of squads working in
% related areas form a \textit{tribe}. A \textit{chapter} brings people with the
% same expertise together across squads and tribes.

\ING is currently leveraging a major shift in the organization to adopt AI to
improve its services and increase business value. As part of it, \ING is defining standards for the different processes around the lifecycle of Machine Learning applications.
The challenges that \ING is
facing at the moment make it an interesting case for our study and allow us to
identify gaps between current challenges by the industry and
academia.

\subsection{Research Methodology}

Semi-structured interviews are the main source of data in this case study. The
data is later triangulated with other resources available inside the
organization. 
The approach used to collect information from interviews and to report data is
based on the guidelines proposed by Halcomb et al.~\citep{halcomb2006verbatim}. It is
a reflexive, iterative process:

\begin{enumerate}
    \item Audio taping of the interview and concurrent note-taking.
    \item Reflective journaling immediately post-interview.
    \item Listening to the audiotape and revising memos.
    \item Triangulation.
    \item Data analysis.
\end{enumerate}

\subsubsection{Participants}
We selected interviewees based on their role and their involvement in the
process of developing Machine Learning applications. We strove to include
people of many different roles and from many different departments. The starting
position for finding interviewees was the lead of a Software Analytics research
team within \ING. More interviewees were found by
the recommendations of other interviewees. The interviewees were also able to suggest
other sources of evidence that might be relevant. We increased the number
of participants until we reached a level of saturation in the remarks mentioned
by interviewees for each stage of the lifecycle.

We adopt a basic approach to assess data saturation. We assume that we achieve data saturation when practitioners from different teams cease bringing insights that we have not observed in previous interviews. Moreover, we only stop collecting data after having data saturation with three consecutive participants.

In total, we interviewed seventeen participants. An overview of the selected participants, with
their role and department, can be seen in Table~\ref{tab:interviewees}. The sixth interview
involved two participants. Therefore, they are labeled as P06a and P06b.

\begin{table}
\caption{Overview of Interviewees}
\label{tab:interviewees}
\begin{tabular}{l l l}
    \hline
    ID & Role & Department\\
    \hline
    P01 & IT Engineer           & Application Platforms           \\
    P02 & IT Engineer           & IT Infrastructure Monitoring    \\
    P03 & Productmanager        & Financial Crime                 \\
    P04 & IT Architect          & Enterprise Architects           \\
    P05 & IT Engineer           & IT4IT                           \\
    P06a* & Advice Professional & Model Risk Management           \\
    P06b* & Advice Professional & Model Risk Management           \\
    P07 & Manager IT            & Global Engineering Platform     \\
    P08 & Feature Engineer      & Data \& Analytics               \\
    P09 & Data Scientist        & Wholesale Banking Analytics     \\
    P10 & Data Scientist        & Chapter Data Scientists         \\
    P11 & IT Engineer           & Application Platform            \\
    P12 & Data Scientist        & AIOps                           \\
    P13 & Data Scientist        & Wholesale Banking Analytics     \\
    P14 & Data Scientist        & Financial Crime                 \\
    P15 & Data Scientist        & Analytics                       \\
    P16 & Data Scientist        & Chapter Data Scientists         \\
    \hline
    \end{tabular}
   \begin{flushleft}
   \footnotesize\emph{*The sixth interview involved two participants, labeled P06a and P06b.}
   \end{flushleft}
\end{table}

\subsubsection{Interview Design}
\begin{figure}
  \centering \includegraphics[width=.9\textwidth]{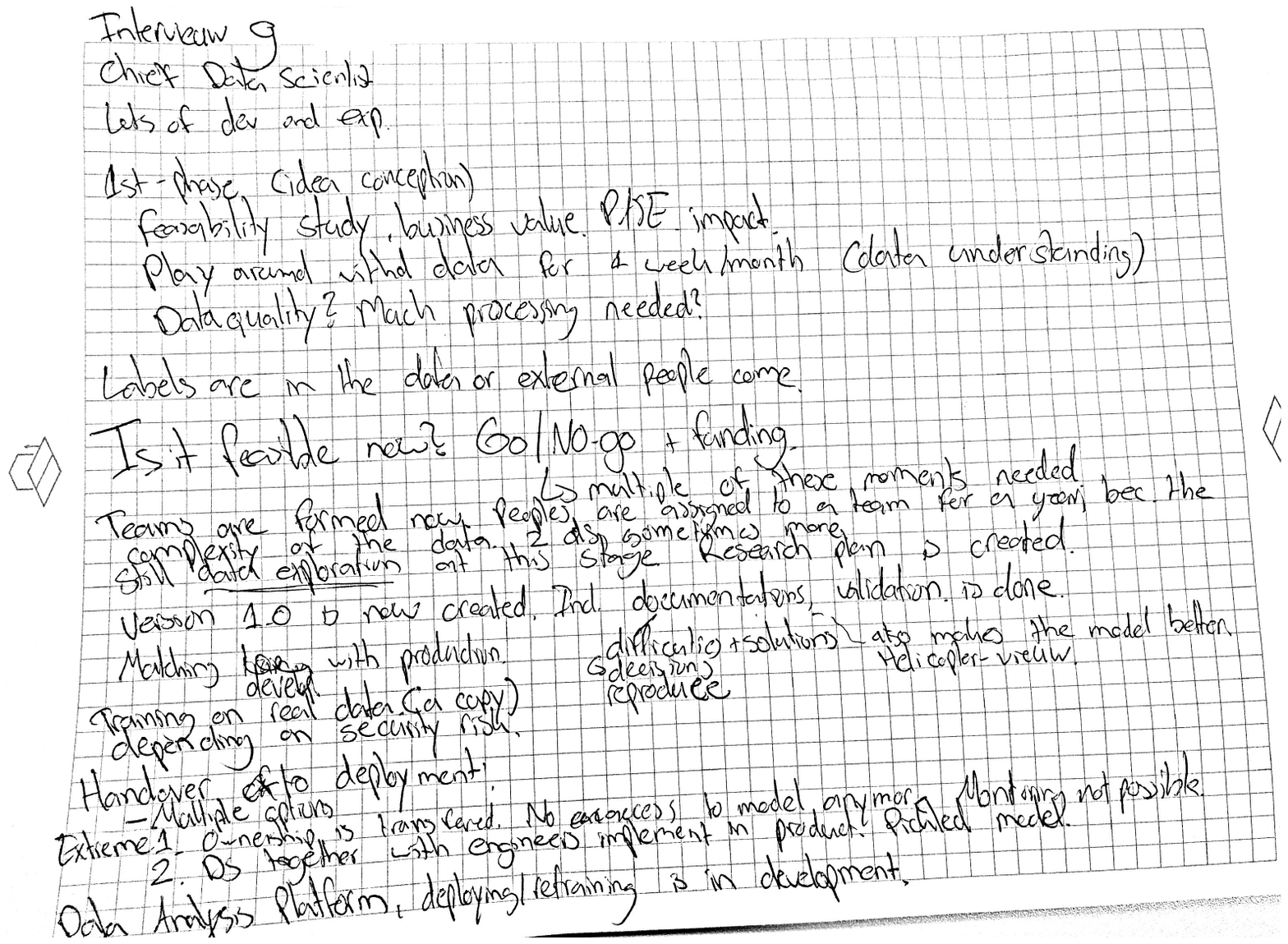}
  \caption{Excerpt of the notes taken during an interview.}
  \label{fig:notes}
\end{figure}

The first two authors conducted the interviews, which took approximately one hour.
We took notes during the interviews and we recorded the interviews with the
permission of the participants. An example of the notes taken with P09 is shown in Fig.~\ref{fig:notes}. This section outlines the main steps of our
interview design. The full details can be found in our corresponding case study
protocol~\citep{haakman2020protocol}.

As interviewers, we started by introducing ourselves and provided a brief
description of the purpose of the interview and how it relates to the research
being undertaken. We asked the interviewees to introduce themselves and
describe their main role within the organization. After the introductions, we
asked the interviewee to think about a specific Machine Learning project he or she was working
on recently. Based on that project, we asked the interviewee to describe all the
different stages of the project. In particular, we asked questions to
understand the main challenges they faced and the solutions they had to
design.

\subsubsection{Post-interview Strategy}\label{sec:post-interview}

\begin{figure}
  \centering
      \vspace{-3ex}
    \includegraphics[width=\textwidth]{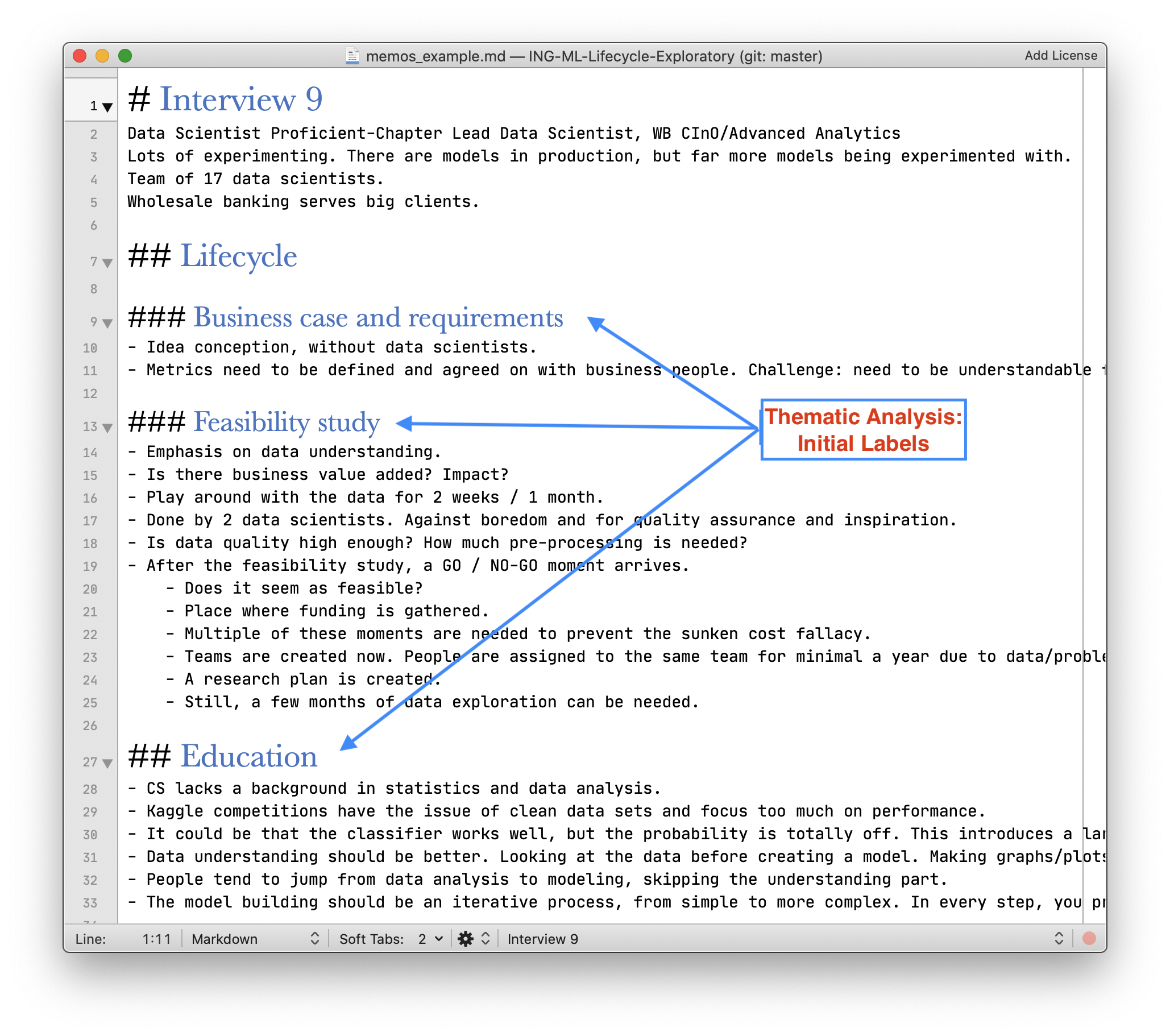}
    \vspace{-6ex}
  \caption{Excerpt of the results of the memoing process. The notes were assigned under different lifecycle themes.}
  \label{fig:memoing}
\end{figure}

Right after each interview, the two interviewers got
together for a collaborative \emph{memoing} process (also called \emph{reflective
journaling}~\citep{halcomb2006verbatim}) combined with thematic coding. Memoing is the review and
formalization of field-notes and expansion of initial impressions of the interaction with more
considered comments and perceptions. Memoing is chosen over creating verbatim transcriptions,
because the costs associated with interview transcription, in terms of time, physical, and human
resources, are significant. An example of the output of memoing is depicted in
Fig.~\ref{fig:memoing}. Also, the process of memoing assisted the researchers to capture their
thoughts and interpretations of the interview data~\citep{wengraf2001qualitative}. The audio
recordings could still be used to facilitate a review of the interviewers' performance, and assist
interviewers to fill in blank spaces in their field notes and check the relationship between the
notes and the actual responses~\citep{fasick1977some}.

The interviewers took between 30--45 minutes to refine their notes. In this process, the notes were
revised and coded into themes based on the particular lifecycle stage that it addressed. We resort
to the \emph{thematic analysis} technique~\citep{fereday2006demonstrating} to derive themes. Thematic
analysis is a qualitative data analysis method divided into four steps: 1) familiarization with
data, 2) generating initial labels, 3) reviewing themes, 4) defining and naming
themes. This technique has been successfully used in previous software engineering
studies to extract patterns from software~\citep{cruz2019catalog}.

The first two authors worked together to discuss and validate the themes. After the first iteration
of label generation -- step 2 of thematic analysis -- we counted with 49 labels. An example of the derived labels is depicted in Fig.~\ref{fig:memoing}. This step was
followed by reviewing themes -- step 3 of thematic analysis -- in which we discussed each label and
looked for other labels that could be redundant. For example, we merged the labels
\emph{Feasibility Study} and \emph{Proof of Concept} together into a single theme. This step
yielded 11 overarching themes that we further detail in Section~\ref{sec:datanalysis}.

Whenever possible, the selected themes follow the nomenclature of
existing frameworks -- namely, CRISP-DM, TDSP, and \citeauthor{amershi2019software}'s model.
Alternative themes were created when 1) we encountered notes that did not fit the existing themes,
or 2) there was a theme being mentioned on different occasions which helped understand a
particular part of the process.

After some time, the interviewers amended the memos by reviewing
the audiotapes. The purpose of this stage was to ensure that the memos provided an
accurate reflection of the interviews~\citep{halcomb2006verbatim}.
Each interview resulted in three artifacts: the recording of the interview, the
field notes taken during the interview, and the memos as a result of the above-mentioned memoing.

\subsubsection{Triangulation}

The main goal of triangulation is to provide means of assessing the validity of insights from practitioners.
We used the documentation in the intranet of \ING to gain a deeper
understanding of the platforms and processes mentioned in the interviews. This documentation is available to all employees in the organization and aims to provide a clear understanding of the processes and resources available. It typically consists of slide decks, short guides, and webpages. This documentation is confidential, thus triangulation was performed by the authors affiliated at \ING.

Ultimately, triangulation did not serve as a mean to discard insights, but rather to understand their relevance and whether they generalize to other sectors of the organization. 
As an example, we observed that, although several teams mentioned being using templates to document machine learning projects, these are not available to the rest of the organization. On contrary, we have analyzed several resources regarding feasibility studies -- meaning that this is a well-established standard at \ING. 
%!TEX root = ../main.tex

\section{Data Analysis} \label{sec:datanalysis}

The input of the interviewees does not answer the research questions directly.
Therefore, we report the resulting data of the interviews in this section and
we use this data to answer the research questions later in
Section~\ref{sec:data_synthesis}.

We organize the data among eight core Machine Learning lifecycle themes: \emph{problem design},
\emph{requirements}, \emph{data engineering}, \emph{modeling}, \emph{documentation}, \emph{model evaluation}, \emph{model deployment}, and
\emph{model monitoring}. Overarching data that does not fit a single lifecycle stage is categorized under
\emph{testing}, \emph{iterative development}, and \emph{education}. In some cases, sub-themes were also defined: \emph{Feasibility study}, \emph{Model Risk Assessment}, \emph{Data Collection}, \emph{Data Understanding}, and \emph{Data Preparation}. These stages were determined based on the thematic coding described in Section~\ref{sec:post-interview}. We refrain from describing details that did not add to existing lifecycle models (e.g., model training).

For all the remarks, we identify the practitioner who mentioned them by
referencing the corresponding ID from Table~\ref{tab:interviewees}. Given that
this is a qualitative analysis, the number of individuals supporting a
particular result has no quantitative meaning on its relevance. The end of each category provides a highlight box with a summary of the main results.

\subsection{Problem Design} \label{sec:problem-design}
Machine Learning projects at \ING start with the definition of the problem that needs to be solved.
Two main approaches are observed in this study:

\begin{enumerate}
  \item Innovation push: a stakeholder comes up with a question or problem that
  needs to be solved. A team is set up to design a solution using a suitable Machine Learning
  technique.

  \item Technology push: a team identifies new data or a set of Machine Learning techniques
  that may add business value and are potentially useful or solving problems within
  the organization. This approach aims to optimize processes, reduce manual
  work, increase model performance, and create new business opportunities.
\end{enumerate}

The problem is defined together with stakeholders and it is assessed whether using Machine Learning is
appropriate to solve the problem~(P01, P14, P15). In the teams of P15 and P14,
this is done by collaboratively filling in a project document with the stakeholders which
contains information like the problem statement, goals, and the corresponding
business case. Also, domain experts outside the teams are part of this.

\highlight{Machine Learning projects start with a problem statement which is used to discuss whether a Machine Learning solution is necessary. This step requires high engagement from problem domain experts.}

\subsection{Requirements} \label{sec:requirements}
% This section should not list the common requirements, but elaborate how the stage looks like.
Besides project-specific requirements, many of the requirements come from the
organization and are applicable to every Machine Learning application~(P15).
These requirements include traceability, interpretability, and
explainability~(P01, P04, P07, P15). Together with all other regulatory
requirements, they pose a big challenge while developing Machine Learning
applications~(P04). A natural consequence of regulatory requirements is that
black-box AI models cannot be used in most
situations~(P01, P04, P14). For risk management safeness, only
interpretable/explainable AI models are accepted.

Project-specific requirements are often defined by the product owner together
with the stakeholders~(P10). Data
requirements are said to become more clear while working with the model~(P04).
As the users of the system are often no Machine Learning experts, defining the model performance
requirements is sometimes a challenge~(P09, P13).

% From Zhang and Harman:
% functional requirements (i.e., correctness and model relevance) and non-functional requirements (i.e., efficiency, robustness3, fairness, interpretability).

\highlight{Requirements are not always defined beforehand. Data and Model requirements become more clear while working with an initial model. Requirements related to traceability, interpretability, and explainability are typically defined a the organizational level.}

\subsection{Data Engineering}
Interviewees describe that data engineering requires the major part of the
lifetime of a Machine Learning project~(P03, P10, P15) and is also the most important for the
success of the project~(P10).

\subsubsection{Data Collection} \label{sec:datacollection}
Data collection is considered a very challenging and time-consuming task~(P03,
P04, P12, P14). Typical use cases require access to sensitive data, which
needs to be formally requested. \ING has an extensive data governance framework
that, among others, assigns data management roles (e.g. data owner) and rules for
obtaining, sharing, and using data. Each dataset is assigned a criticality
rating, to define the degree of data governance and control required.

There might be people with different access privileges to data in the same project.
This means that, in the exploratory stages of some projects using critical data,
only a restricted number of team members (e.g., data scientists) are able to
perform an exploratory analysis of data. The remaining practitioners will only
have access to the model specification~(P04).

A challenge of data collection is making sure that the (training and test) data collected is
representative of the problem~(P13). As an example, if a Machine Learning model is trained on systems logs,
it should be made sure that logs of all systems are available.
Another challenge is merging data from multiple sources~(P10, P12). Going back to the logging example,
different systems may have different logging formats, but the configurations of these
formats cannot be altered by the developers creating the model.

\subsubsection{Data Understanding}
In the data understanding stage, an assessment is done on the quality of available data and how
much processing will be required to use that data. It comprises exploratory data
analysis, often including graphical visualizations and summarization of data.
According to P09, the temptation of applying groundbreaking Machine Learning techniques tends
to overlook the importance of understanding the data.

Data understanding is also an important step to assess the feasibility of the
project. Thus, it entails not only performing an exploratory analysis, but also
a considerable effort in communicating the main findings to all the different
stakeholders. This tends to be a slow process~(P12).

\subsubsection{Data Preparation}
After the data is collected and it is assessed that the data is representative
of the problem being solved, the data is prepared to be used for modeling.

A challenge regarding data preparation is that the same
pre-processing has to be ensured in the development environment and in the
production environment~(P08, P09). Data streams in production
are different than in the development environment and it is easier to clean
training and testing data than production data~(P09).

\highlight{Collecting, understanding, and preparing data are the most time-consuming stages of
Machine Learning projects. There is a meticulous data access control that, despite being quintessential, sets major obstacles in understanding the data and performing exploratory analyses. Practitioners emphasize, data understanding implies being able to communicate it to other stakeholders. Finally, the differences between development and production environments pose challenges for data preparation.}

\subsection{Modeling}\label{sec:modeling}
Model training is mostly done in on-premises environments such as
Hadoop\footnote{Hadoop enables distributed processing of large data sets across
clusters of computers \Website{https://hadoop.apache.org}} and
Spark\footnote{Spark is a unified analytics engine for large-scale data
processing. \website{https://spark.apache.org}} clusters~(P09) or in generic
systems using, for example, the scikit-learn\footnote{Scikit-learn is a Machine Learning library for Python. \website{https://scikit-learn.org}} library~(P01).
These private platforms are connected with the data lakes where data is stored,
so training can be done on (a copy of) real production data~(P01, P03).
The on-premises environment has
no outgoing connection to the internet, so a connection to other cloud
services such as Microsoft Azure\footnote{Microsoft Azure is a cloud computing
service. \website{https://azure.microsoft.com/en-us}} or Google
Cloud\footnote{Google Cloud is a cloud computing service.
\website{https://cloud.google.com}} is not possible~(P08). This
means that data scientists are limited to the tools and platforms available
within the organization when dealing with sensitive data. Also, all project dependencies need to be previously approved, after which they are made available in a private
package repository~(P04, P12), which contains whitelisted packages
that have been internally audited. This can be frustrating, when new ground-breaking AI technologies appear, practitioners have to wait before they can explore the potential of those technologies at \ING~(P12) -- we later refer to this challenge as \emph{Technology Access} (cf. Section~\ref{sec:data_synthesis}).
Fewer restrictions are in place if Machine Learning is applied to public data, for example on
stock prices. In that case, external cloud services and packages may be used
(P09).

Model training is an iterative process. Usually, multiple models are created
for the same problem. First, a simple model is created (e.g., a linear
regression model) to set as a baseline~(P09). In the following iterations, more
advanced models are compared to this baseline model. If an approach other than
Machine Learning already exists (e.g., rule-based software), the models are
also compared with this.

To keep track of different versions of models,
different teams use different strategies. For example, the team of P08 keeps
track of an experiment log using a spreadsheet, in which the training set, validation set, model, and
pre-processing steps are specified for each version. This approach for
versioning is preferred over solutions like
MLFlow\footnote{MLFlow is a platform to manage the Machine Learning lifecycle. \Website{https://mlflow.org}} for the sake of simplicity~(P08, P15).

\subsubsection{Model Scoring}
An implicit sub stage of modeling is assessing model performance to
measure how well the predictions of the model represent ground truth data.

We define \emph{Model Scoring} as assessing the performance of the model based
on scoring metrics (e.g., f1-score for supervised learning). It is also known
as \emph{Validation} by the Machine Learning community, which should not be
confused with the definition by the Software Engineering
community\footnote{\emph{Validation} in Software Engineering ``is the set of activities ensuring and gaining confidence that a system is able to accomplish its intended use, goals and objectives''~\citep{iso2015systems}.}~\citep{ryan2017use,iso2015systems}.

The main remarks for this stage are related to defining the right set of
metrics~(P03, P06, P12, P14, P15, P16). The problem is two-fold: 1) identify the
right metrics and 2) communicate why the selected metrics are right.
Practitioners report that this is very problem-specific. Thus, it requires a
good understanding of the business, data, and learning algorithms being used.
From an organization's point of view, these different perspectives are a big
barrier to defining validation standards.

\highlight{The challenges in \emph{Modeling} summarize as follows: 1) the latest Machine Learning technologies are not always eligible for use; 2) baseline models are essential artifacts for model development; 3) teams keep track of all experiments, which often revolves around keeping a customized spreadsheet; and 4) defining performance metrics is problem-specific, posing a challenge to the definition of standards at the organizational level.}

\subsection{Documentation} \label{sec:documentation}
Each model has to be documented~(P02). This serves multiple goals. It makes
assessing the model from a regulatory perspective possible~(P09, P13), it
enables reproducibility, and also can make the model better because it is looked
at from a broad perspective -- i.e., a ``helicopter view'' (P09). It also
provides an audit trail of actions, decisions, versions, etc. that supports
evidencing. Documentation also supports the transfer of knowledge, for example,
to new team members or the end-users which are mostly not Machine Learning experts (P12). Just
like code, documentation is also peer-reviewed~(P13).

The content of the documentation differs slightly per department, but all
documentation should at least follow the minimum standards defined by the model
risk management framework~(P06). Some teams extend on this by creating templates
for documentation themselves~(P13). In general, the following is documented when
developing a Machine Learning application: the purpose, methodology, assumptions,
limitations, and the use of the model. More concretely, a Technical Model
Document is created which includes the model methodology, input, output,
performance metrics and measurements, and testing strategy~(P14). It furthermore states all
faced difficulties and their solutions, plus the main (technical) decisions
(P09). It has to explain why a certain model is chosen and what its inner workings are,
to be able to demonstrate the application
does what the creators claim it is doing. Creating documentation is considered overly time-consuming, although necessary~(P07).

\highlight{
Documentation is a first-class artifact for regulatory compliance, knowledge transfer, and reproducibility. Hence, a peer-review process is in place to ensure documentation quality. 
}

\subsection{Model Evaluation} \label{sec:evaluation}

An essential step in the evaluation of the model is communicating how well the
model performs according to the defined metrics. It is about demonstrating that
the model meets business and regulatory needs and assessing the design of the
model. One key difference between the metrics used in this step and the metrics
used for \emph{Model Scoring} is that these metrics are communicated to
different stakeholders that do not necessarily have a Machine Learning or data
science background. Thus, the set of metrics needs to be extended to a general
audience. One complementary strategy used by practitioners is having live demos
of the model with business stakeholders~(P03, P15, P16). These demos allow
stakeholders to try out different inputs and try corner cases.

\subsubsection{Model Risk Assessment} \label{sec:risk-management}
An important aspect of evaluating a model at \ING is making sure it complies
with regulations, ethics, and organizational values~(P15, P06). This is a common
task for any type of model built within the organization -- i.e., not only Machine Learning
models but also economic models, statistical forecasting models, and so on. In
the interviews, \emph{Model Risk Assessment} was mentioned as mandatory
within the model governance strategy, undertaken in collaboration with an independent specialized team~(P06, P14). This is a long-stablished stage which is now being challenged by the specifics of Machine Learning. For example, traditional risk assessment teams did not initially have the right Machine Learning expertise to evaluate the models with confidence.

Depending on
the criticality level of the model, the intensity of the review may vary.
Each model owner is responsible for the risk management of their model, but
colleagues from the risk department help and challenge the model owner in this
process.

During the periodic risk assessment process, assessors inspect the documentation provided by the
Machine Learning team to assess whether all regulations and minimum standards are followed. The
documentation used in this stage is considered to be overly time-consuming, as emphasized by P07:
``70\% percent of the time people are writing Word documents to explain their code is compliant.''.
Although the process is still under development within \ING, the following key points are being
covered~(P06): 1) model identification (identify if the candidate is a model which needs risk
management), 2) model boundaries (define which components are part of the model), 3) model
categorization (categorize the model into the group of models with a comparable nature, e.g.
anti-money-laundering), 4) model classification (classify the model into in the class of models
which require a comparable level of model risk management), and 5) assess the model by a number of
sources of risk.

\highlight{Although Model Risk Assessment is not new to the fintech industry, Machine Learning is requiring a revised approach. Currently, developers endure considerable efforts to create the required documentation.}

\subsection{Model Deployment}
% Some teams resort to a checklist of technical requirements that need to be met
% before the model can go to production (P15, P16). These technical requirements
% are ensured collaboratively by data scientists and software engineers.

We observed three deployment patterns at \ING:
\begin{enumerate}
  \item A specialized team creates a prototype with a validated methodology, and
  an engineering team takes care of reimplementing it in a scalable,
  ready-to-deploy fashion. In some cases, this is a necessity due to the
  technical requirements of the model, e.g., when models are developed in
  Python, but should be deployed in Java~(P08, P09, P13).

  \item A specialized team creates a model and exports its configuration (e.g.,
  a \emph{pickle}\footnote{A \emph{pickle} is a serialized Python object.
  \Website{https://docs.python.org/3/library/pickle.html}} and required
  dependencies) to a system that will semi-automatically bundle it and deploy it
  without changing the model~(P01, P09).

  \item The same team takes care of creating the model and taking it into production. This mostly means that software engineers are part of the team and a structured and strict software architecture is ensured.
\end{enumerate}

Similar to the training environments, Machine Learning systems are deployed to
on-premises environments. A reported challenge regarding the deployment
environment is that different hardware and platform parameters (e.g., Spark
parameters) can result in different model behavior or errors~(P16). For
example, the deployment environment may have less memory than the training
environment. Furthermore, the resources for a Machine Learning system are
dynamically allocated whenever needed. However, it is not trivial understanding
when a system is no longer needed and should be scaled down to zero~(P01).

\highlight{There are deployment patterns in which a separate team needs to reimplement the model to
meet production settings.}

\subsection{Model Monitoring}\label{sec:results_monitor}
After having a model in production, it is necessary to keep track of its
behavior to make sure it operates as expected. It implies testing the model
while the model is deployed online. The main advantage is that it uses real
data. Previous work refers to this stage as \emph{online
testing}~\citep{zhang2020machine}.

The inputs and outputs of the model are monitored
while it is executing. Each model requires a different approach and different
metrics, as standards are not yet defined. In this stage, practitioners also
look into whether the statistical properties of the
target variable do not change in unforeseen ways~(P11). The model behavior is
mostly monitored by data science teams and is still lacking automation~(P03,
P05, P06, P14). Also, the impact on user experience is monitored when the model has a
direct impact on users. This is mostly done using A/B testing
techniques and can have business stakeholders directly involved~(P03, P10).

Teams resort to self-developed or highly-customized dashboard platforms to
monitor the models~(P15, P16). Within the organization, different teams may have
different platforms. While standardization is in development, for now, we have
not observed solutions that are used across the
organization. A big challenge in making these platforms available is the fact
that each problem has different monitoring requirements and considerable
engineering efforts need to be undertaken to effectively monitor a given model
and implement access privileges~(P15).

\highlight{More automation is needed for model monitoring. Teams have created their own automation tools, but making them available to other teams requires unfeasible efforts that do not meet their priorities. }

\subsection{Testing}
Testing is a task that is transversal to the whole development process. It is done at
the model level and at the software level.

Testing at the model level addresses requirements such as correctness, security,
fairness, and interpretability. With the exception of correctness, we have not
observed automated approaches to verify these requirements. A challenge for the
correctness tests is defining the number of errors that are acceptable -- i.e., the right
threshold~(P14).

For testing at the software level, unit and integration testing is the general
approach. It scopes any software used in the lifecycle of the model~(P07). It
enables the verification of whether the techniques adopted in the design of the
Machine Learning system are working as expected. However, although unit and integration
testing is part of the checklist used for \emph{Model Evaluation}, a number of
projects are yet not doing it systematically~(P12, P15). As reported by P14,
tests are not always part of the skill set of a data scientist. Nevertheless,
there is a generalized interest in learning code testing best practices~(P12).

\highlight{Although practitioners are eager to learn automated testing practices, this is not part of their skillset. Hence, projects are struggling to adopt unit and integration testing strategies. }

\subsection{Iterative Development} \label{sec:resuls-agile}
At \ING, teams adopt agile methodologies. Three practitioners~(P03, P09, P14)
mentioned that using agile methodologies is not straightforward in the early
phases of Machine Learning projects. They argued that performing a feasibility study does not
fit in small iterations. The first sprint requires spending a considerable
amount of time understanding and preparing data before
being able to deliver any model. On the other hand, interviewees acknowledge the
benefits of using agile~(P03, P14). It helps keep the team focused on practical
achievements and goals. Another advantage is that stakeholders are kept in the loop~(P14).

Typically, 2--3 data scientists are working together on the same model. For this
reason, issues with having many developers working on the same model and merging
different versions of a model have not been disruptive yet.

\subsubsection{Feasibility Study}
The end of the first iteration is also a decisive step in the project. Based on
the outcome of this iteration there is a go/no-go assessment with all the
stakeholders, in which the project is evaluated in terms of \emph{viability} (i.e.,
does it solve a business issue), \emph{desirability} (i.e., is it complying with
ethics or governance rules), and \emph{feasibility} (i.e.,
cost-effectiveness)~(P04, P09, P15, P16). This process is well-defined within
the organization for all innovation projects.
According to P04 and P09, feasibility assessments are essential at any
point of the project -- it is important to adopt a \emph{fail-fast} approach.

\highlight{All projects must go over a feasibility study in their early stages. Until then, projects do not fit the typical sprint-based agile planning. An agile approach helps practitioners prioritize tasks and engage stakeholders.}

\subsection{Education}
Interviewees indicated multiple ways in which education can be improved to make
graduates better Machine Learning practitioners in the industry. Firstly, data scientists
should have more knowledge of Software Engineering and vice-versa~(P01, P11,
P14, P16). P11 indicates that data scientists with little software
engineering knowledge will produce code that is harder to maintain and likely increases
technical debt. On the other hand, a software engineer without data science
expertise may write clean code, which nevertheless may not add much business value,
because of ineffective data exploration strategies~(P09).

Another remark by practitioners is that education should focus more on the process of developing
Machine Learning, rather than teaching learning techniques~(P08). While graduates are appreciated
for their broad sense of the state-of-the-art, they must learn how to tackle Machine Learning
issues in large organizations~(P08, P10). Academia knows well how to work with new projects, but in
reality, the history of the company affects how to perform Machine Learning -- e.g., integration
with legacy systems~(P08). Graduates seem to underestimate the efforts needed for data engineering,
especially data collection~(P03, P09, P12). Also, too much attention lies solely on the performance
of models. In reality, over-complex models cannot be applied in organizations, because they tend to
be too slow or too hard to explain~(P16). These models -- squeezing every bit of performance -- are
great for data science competitions as facilitated on Kaggle, but not for the industry, where more
efficient solutions are necessary~(P09, P16).

\highlight{There is practical value on having a strong background on both Software Engineering and
Data Science. Education should put more focus on the process instead of model-training techniques.}

%!TEX root = ../main.tex

\section{Data Synthesis} \label{sec:data_synthesis}

In this section, we answer each research question.

\rquestionbox{rq:lifecycle} % How do existing Machine Learning lifecycle models fit the fintech domain?

Our interviews show evidence that existing models do not fit the needs of the today's fintech industry and changes ought to be made.

To explain this further, we pinpoint the differences between lifecycle models
existing in the literature and the findings observed in our study. We select
three reference models, as described in Section~\ref{sec:existing}:
\emph{CRISP-DM}~\citep{shearer2000crisp}, \emph{TDSP}~\citep{TDSP}, and \citet{amershi2019software}. 
We justify and define each required change and discuss the constraints to which they generalize outside the case of \ING{}: to the
fintech domain or to general Machine Learning projects.

\begin{figure}
  \centering \includegraphics[width=0.6\textwidth]{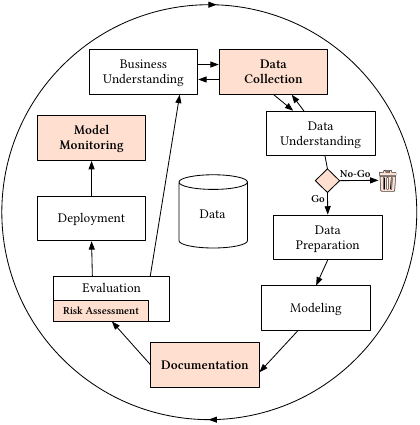}
  \caption{Refined CRISP-DM model. Additions in red, with bold text.}
  \label{fig:crisp-dm-improved}
\end{figure}

We propose the changes of
CRISP-DM in Fig.~\ref{fig:crisp-dm-improved} -- new stages are highlighted with orange background and bold text. We add three new essential
stages: \emph{Data Collection} (as part of \emph{Data Engineering}),
\emph{Documentation}, and \emph{Model Monitoring}. Furthermore, we emphasize the
feasibility assessment with the ``Go/No-go'' checkpoint and a sub-stage
\emph{Model Risk Assessment}, part of \emph{Evaluation}.

There are, however, stages identified at \ING that naturally fit CRISP-DM. Similarities between
CRISP-DM and the stages observed are \emph{Business Understanding}, \emph{Data Understanding},
\emph{Data Preparation}, \emph{Modeling}, \emph{Evaluation}, \emph{Deployment}.

\begin{figure}
  \centering \includegraphics[width=0.8\textwidth]{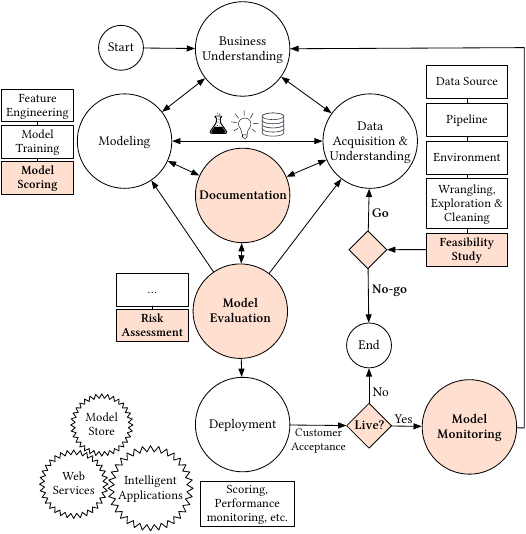}
  \caption{Refined TDSP model. Additions in red, with bold text.} \label{fig:tdsp-improved}
\end{figure}

As depicted in Fig.~\ref{fig:tdsp-improved}, we adapt the TDSP model to
include \emph{Documentation}, \emph{Model Evaluation}, and \emph{Model
Monitoring} as major stages. We also emphasize \emph{Model Risk Assessment} (as part
of \emph{Evaluation}) and a \emph{Feasibility Study}.

We observed stages that are already being covered by TDSP: \emph{Business
Understanding}, \emph{Data Acquisition \& Understanding}, \emph{Modeling}, and
\emph{Deployment}.

\begin{figure}
  \centering \includegraphics[width=1.0\textwidth]{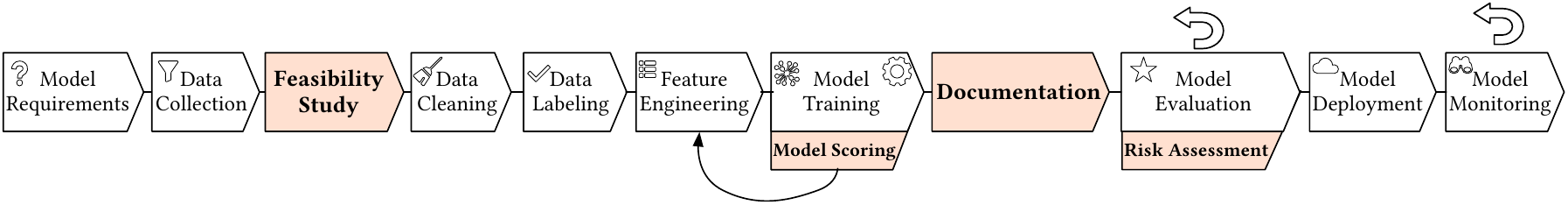}
  \caption{Refinements to the Microsoft model described by \citet{amershi2019software}. Additions in red, with bold text.} \label{fig:amershi-improved}
\end{figure}

After inspecting the model by \citet{amershi2019software}, we propose the changes in Fig.~\ref{fig:amershi-improved}. We adapt the original model to include \emph{Feasibility Study}, and \emph{Peer-reviewed Documentation}. We also emphasize \emph{Model Scoring} (as part of \emph{Model Training}) and \emph{Risk Assessment} (as part of Model Evaluation).
Other observations in our study naturally fit the stages described by \citet{amershi2019software}: \emph{Model Requirements}, \emph{Data Collection}, \emph{Data Cleaning}, \emph{Data Labeling}, \emph{Feature Engineering}, \emph{Model Training}, \emph{Model Evaluation}, \emph{Model Deployment}, \emph{Model Monitoring}.

There are, however, stages identified at \ING that naturally fit CRISP-DM and TDSP. 
Similarities between TDSP are \emph{Business
Understanding}, \emph{Data Acquisition \& Understanding}, \emph{Modeling}, and
\emph{Deployment}.

The adaptations of the models will be further elaborated upon in the following
paragraphs.

\paragraph{Data Collection} Although CRISP-DM encompasses \emph{Data Collection}
within \emph{Data Understanding} and \emph{Data Preparation}, our observations
reveal important tasks and challenges that need to be highlighted. As reported
in Section~\ref{sec:datacollection}, \emph{Data Collection} requires getting
privileges to access data with different criticality-levels and making sure the
data is representative of the problem being tackled. Our proposition is that the
characteristics observed at \ING regarding this phase generalize to any large
organization dealing with sensitive data. TDSP and \citeauthor{amershi2019software} already contemplate this stage.

\paragraph{Go/No-go or Feasibility Study} The aforementioned \emph{Feasibility
study} (cf. Section~\ref{sec:resuls-agile}) is an essential part of a Machine
Learning project to ensure projects have everything in place to deliver the
long-term expectations. It was a recurrent step observed in our study, which is
aligned with the agile approach, \emph{Fail Fast}, promoted at \ING{} and many
organizations alike. It may generalize to other cases, depending on the agile
culture of the organization.

\paragraph{Documentation} In our case, documentation revealed to be a quintessential artifact for a
Machine Learning project. Documentation is the key source of knowledge on how the model is
designed, evaluated, tested, deployed, and so on. The documentation is used to evaluate, maintain,
debug, and keep track of any other decision regarding the model. It is hard to replace
documentation with other strategies because stakeholders with a non-technical background also need
to understand the model and have confidence in how the Machine Learning model is designed. Although
documentation is also important in traditional Software Engineering applications, the codebase is
usually the main target of analysis from audits. In Machine Learning, documentation contains
important problem-specific decisions that cannot be understood in the code itself. Hence, at
\ING{}, Machine Learning practitioners devote a big part of their time to write clear
documentation. Furthermore, documentation endures a peer-review process to make sure it is sound
and complete. We have no evidence on how this stage generalizes to other organizations, but believe
this to be crucial in any highly regulated environment.

\paragraph{Model Evaluation} Although the original version of TDSP also included \emph{Model
Evaluation}, it was proposed as an activity under the \emph{Modeling} stage. The same applies to
the model by \citeauthor{amershi2019software} that describes \emph{Model Evaluation} as a stage
where ``the engineers evaluate the output model on tested or safeguard datasets using pre-defined
metric''. For this particular purpose, we use the term \emph{Model Scoring} and consider it a part
of the \emph{Modeling} or \emph{Model Training} activities. However, there is an important part of
the evaluation that requires more stable versions of the models. Moreover, it is undertaken with
stakeholders that are not part of the \emph{Modeling} loop -- e.g., live demos with business
managers (cf. Section~\ref{sec:evaluation}) -- and requires assessing the model in terms of
business metrics. Thus, we highlight this part of the evaluation as its own stage. An important
part of this stage is model risk assessment, which we further explain below.

\paragraph{Model Risk Assessment} Model Risk Assessment is crucial to any banking or finance
organization. Although companies in the finance domain already have a big history of traditional
risk management, it does not consider Machine Learning models. Moreover, risk assessment teams do
not necessarily have a Machine Learning background to make an informed decision. This does not mean
that Machine Learning models face a less meticulous risk assessment -- it rather means that the
process will take even more time, being a bottleneck in the lifecycle. Thus, at \ING{}, model risk
assessment is enduring a major transformation to make sure this is not a bottleneck in the process.
Notwithstanding its importance, none of the studied lifecycle models contemplate this stage.

\paragraph{Model Monitoring} Most Machine Learning models operate continuously and produce outputs
online. Our study shows that the natural step after deployment is \emph{Monitoring} -- for example,
using dashboards -- to ensure the model is behaving as expected. \emph{Model Monitoring} is not
explicit in neither CRISP-DM nor TDSP, but it is relevant to any domain. In fact,
\citeauthor{amershi2019software} already contemplates this stage.

Finally, although not depicted in the proposed lifecycles, \emph{Education} is
a stage implicit throughout the whole lifecycle. We observe that universities
and courses on Machine Learning need to provide a more holistic approach to
focus on all the different stages of the lifecycle of a Machine Learning system.

A lifecycle stage that we did not yet observe is the end of life of a Machine
Learning system -- i.e., the \emph{Disposal} stage. We presume that a disposal stage is
not relevant yet due to the recency of Machine Learning in fintech.

\rquestionbox{rq:challenges} % What are the specific challenges of developing Machine Learning applications in fintech organizations?

We highlighted many challenges of developing Machine Learning applications in
Section~\ref{sec:datanalysis}. While most challenges potentially affect any tech-company leading an AI-powered digital transformation, there are two that stand out in the fintech domain: \emph{Model Governance} and \emph{Technology Access}.

\emph{Model Governance} is on top of the agenda of the case in this study. A
well-defined process is in place to validate regulations, ethics, and social
responsibility in every Machine Learning model. The relevance of this problem
to fintech organizations goes beyond Machine Learning applications: math-based
financial models have long been deployed under well-defined risk management
processes.

Nevertheless, we observe a need to revise and recreate model governance that suits the
particularities of models that are now automatically trained. E.g., continuous training -- a
practice that is essential for any high maturity Machine Learning
process~\citep{akkiraju2020characterizing,lwakatare2020data} -- does not fit the traditional model
risk assessment approach in fintech. A new set of documentation, and a manual audit are the bare
minimum to release a new version of the model. Hence, automated tools for model governance are
essential to ensure Machine Learning models comply with regulations and reduce bottlenecks in the
development process.

Our results imply that RegTech -- the branch of fintech for managing regulatory requirements -- is
an emergent field with direct contributions to intelligent systems in fintech. Having automated
mechanisms to check model compliance with regulations is essential for the adoption of continuous
integration for AI systems in fintech companies. This is an important challenge since, according to
previous work~\citep{serban2020adoption}, continuous integration is perceived by practitioners as
one of the unexploited practices with the most potential.

Moreover, model risk experts are now required to have a strong background in two disjoints fields: 1) \emph{Governance, Risk Management, and Compliance} and 2) \emph{AI}. We conjecture that this challenge generalizes to other heavy-regulated domains, such as \emph{LegalTech} and \emph{Healthcare Technology}.

\emph{Technology Access} All AI technologies, tools, and libraries need to be audited to make sure
they are safe to be used in fintech applications. Only then, practitioners are able to design their
Machine Learning systems around the latest technology. This is a challenge that needs to tackled by
any organization akin to \ING. As presented in Section~\ref{sec:modeling}, this process can be limiting since new AI technologies are appearing every day. Practitioners willing to try the latest AI technology may feel less motivated since it may take some time before they are approved.
As referred in Section~\ref{sec:problem-design}, many problems at \ING are triggered by the \emph{Technology push}. Hence, new business opportunities might be missed if practitioners are not able to experiment the latest AI technologies.

We do not know to what extent \emph{Technology Access} is also a challenge to software
organizations operating in other domains. Previous work suggests that only 8\% of software
developers consider an organization's culture and policies highly-influential when selecting
third-party software libraries~\citep{larios2020selecting}. On contrary, 52\% consider it as a
low/moderate influential factor. Nevertheless, we argue that similar obstacles might be observed in
many other organizations with high-maturity software development processes. Hence, industries that
want to shift towards AI-based systems need to be able to quickly, yet safely, adopt new
technologies.

%!TEX root = ../main.tex

\section{Discussion} \label{sec:discussion}

In this section, we discuss the implications of our results and elaborate on the threats to the validity of our findings.

\subsection{Implications} \label{subsec:implications}
We see the following implications of our results for the fintech industry and
for research.

\subsubsection{Implications for Machine Learning Practitioners}
% Arie: be aware of extra steps and challenges.
Machine Learning practitioners have to be aware of extra steps and challenges in
their process of developing Machine Learning applications. Although not
mentioned in existing lifecycle models, the undertaking of feasibility
assessments, documentation, and model monitoring, are crucial while developing
Machine Learning applications.

\subsubsection{Implications for Process Architects}
% Arie: update your models as we did, include this in your training,
Existing lifecycle models provide a canonical overview of the multiple stages in
the lifecycle of a Machine Learning application. However, when being applied to
a particular context, such as fintech, these models need to be adapted. From our
findings, we suspect that this is also the case for other fields where AI is
getting increasing importance. Process architects for intelligent systems for
healthcare, autonomous driving, among many others, need to look at their
lifecycle models from a critical perspective and update the models accordingly.

Moreover, process architects ought to keep an eye on the Technology Access blocker. Our work suggests that the process of approving the latest AI technologies should be ahead of the needs of practitioners. We argue that a pro-active process should be in place to audit AI technologies. It is a key factor to explore new business opportunities and to keep developers motivated.

\subsubsection{Implications for Researchers}
% Arie: tool support, additional domains, ...
Researchers could focus on solving the reported challenges in the Machine
Learning lifecycle with additional tool support and reveal challenges of the ML
lifecycle in other domains by extending the case study to more organizations and
different types of industries.

More automation is required for exploratory data analysis and data integration
techniques~\citep{mitchell2019model,damiani2018towards}. Moreover, there are minimal advancements in
documentation of Machine Learning projects. Techniques ought to be studied to help trace
documentation back to the codebase and vice versa.

Furthermore, solutions to challenges in the ML lifecycle should be researched. Our study shows
that, despite the increasing trend on improving the state-of-the-art model training techniques,
there is a research gap on the challenges of developing real-world machine learning systems. For example, our work shows that it is important to move from \textbf{model-centric AI} towards \textbf{data-centric AI}\footnote{Andrew Ng nicely explains the importance of data-centric AI in his webinar ``A Chat with Andrew on MLOps: From Model-centric to Data-centric AI''. Recording available online, retrieved on \today: \url{https://www.youtube.com/watch?v=06-AZXmwHjo}}. Practitioners spend most of their time collecting and understanding data, rather than training the model per se. Ultimately, the success of a model stems from the quality of the training and test data, which is where practitioners spend most of their efforts. Moreover, to the best of our knowledge, there is no research literature addressing the challenges of auditing and approving AI software libraries in large-scale organizations. 

More research should focus on assisting model governance to reduce bottlenecks in the development
process and help ensure that Machine Learning models comply with regulations. Ongoing work argues
that model governance literature for fintech is wide and lacks a coherent research
agenda~\citep{kavuri2019fintech}. Yet, related literature suggests that the problem ought to be
addressed not only by the fintech industry, but also from the perspective of regulators who have to
adapt~\citep{brummer2018fintech,van2018making}.

\subsubsection{Implications for Tool Developers}

Although a number of tools are emerging to aid ML engineering, these solutions
fail to address the singularities of different projects. Thus, practitioners
are adopting their own customized solutions. For example, spreadsheets are
still being used to manually log experiments regardless of the existing
automated solutions, such as MLFlow, DVC, Replicate, and so on. It is important
to understand what is missing in the current solutions and how we can propose a
solution that effectively solves version control to keep track of changes in
data, changes in scoring metrics, and executions of different experiments.

Software testing needs to be extended and adapted for Machine Learning software to help effectively
test the Machine Learning pipeline at software-, data-, and model-level. It is also necessary to
create holistic monitoring solutions that can scale to different models in an organization. There
is a need for strategies to help practitioners select the right set of model scoring metrics.
Finally, agile development practices are perceived as beneficial but need to be adjusted for AI
projects.

\subsubsection{Implications for Educators}
Education of Machine Learning should focus on the whole lifecycle of Machine
Learning development, including exploratory analysis with a focus on statistics,
data analysis and data visualization. Moreover, practitioners with background on
both data science and software engineering are a valuable resource for
organizations. This emphasizes the importance of a transdisciplinary approach to
AI education~\citep{wang2003cognitive,nicolescu2008transdisciplinary} and it is
congruent with previous work that reports that a Software Engineering mindset
brings more awareness on the maintainability and stability of an AI
project~\citep{arpteg2018software}.

\subsubsection{Implications for Organizations Embracing AI}

The embrace of AI stretches the adequacy of well-established processes at organizations.
Multi-disciplinary teams are essential to embrace AI: AI experts have the knowledge to try
innovative approaches, but will likely have little expertise to identify business value. Thus,
knowledge transfer between stakeholders is challenging and might hinder the motivation of
developers. New strategies must be outlined to reduce the amount of effort required to document AI
projects. Providing AI training to employees can help enable the transition to AI. Not only it
makes discussions and decisions about AI projects more effective, but also it helps to identify
business opportunities in areas that have not explored the potential of AI yet. Finally, it may
happen that different teams will solve the same problem independently (e.g., experiment logs).
Although teams are aware of it, they argue that they do not have enough resources to make their
solutions reusable (cf. Section~\ref{sec:results_monitor}). Thus, organizations should create task
forces to make such tools available to all teams.

\subsection{Threats to Validity} \label{sec:threats}
This subsection describes the threats and limitations of the study design and how
these are mitigated. These limitations are categorized into researcher bias,
respondent bias, interpretive validity, and generalizability, as reported by
Maxwell~\citep{maxwell1992understanding} and
Lincoln et al.~\citep{lincoln1985naturalistic}.

\subsubsection{Researcher Bias}
Researcher bias is the threat that the results of the study are influenced by
the knowledge and assumptions of the researchers, including the influence of the
assumptions of the design, analysis, and sampling strategy.

% Self-selection of participants
%   - Results may be specific to the teams which participated -> interviews until saturation and variety of departments and roles.
A threat is introduced by the fact that participants are self-selected.
This means that there might be employees in the company which should be included
in the study but are not selected. During the planning phase, participants are
selected with different roles and from different departments to have an as
diverse starting point as possible. Thereafter, more participants are found by
the recommendation of other interviewees and employees until we reach
saturation on the information we get from the interviews, i.e. until no new
information or viewpoint is gained from new subjects~\citep{strauss1990basics}.

Moreover, we validated the findings of this study by collecting feedback from relevant stakeholders at ING. Our approach was two-fold: 1) invited relevant stakeholders for a 30-minute presentation of our results, followed by a Q\&A and discussion session, and 2) we sent out a report via email with the results and analysis provided in this study.

The presentation counted with around 15 participants that were not part of the case-study interviews. The main point highlighted by participants was the fact that ING is spending a lot of time and resources to improve their machine learning processes. Hence, they expect to mitigate some of the reported challenges in the meantime.

For the email communications, a total of 19 people were addressed, including stakeholders who have participated in the interviews and stakeholders who have not. One recipient, who had participated in the interviews, took the time to thoroughly read the paper. The recipient mentioned to be in agreement with the findings on the paper but raised one concern -- the fact that there is not yet a standard methodology at ING to develop machine learning systems. Hence, we report our analyses in this paper as critical observations over existing theories, rather than the \emph{de facto} model used at ING.

\subsubsection{Respondent Bias}
Respondent bias refers to the situation where respondents do not provide honest
responses.

% Self-reported data
%   - People tend to sell their project. -> triangulation
%   - People forget to mention things. -> mindmap for relevant topics
%   - People tend to tell how it is supposed to be, in contrast with how it is in reality. -> triangulation and tell he is not being evaluated or judged and ask to think about specific project.
%   - Recording the interviews may affect the responses, participant is extra careful. -> tell recordings will not be published and paper will be checked first.
The results of the interviews rely on self-reported data. All people tend to judge
the past disproportionately positive. This psychological phenomenon is known as
rosy retrospection~\citep{mitchell1997temporal}. Furthermore, interviewees who know golden standards from for example
literature may tell how things are supposed to be, in contrast with how they are
in reality. These biases are mitigated by reassuring interviewees their answers will not be evaluated or
judged and by asking them to think about a particular project they
have been working on.

A methodological choice which can form a threat to validity is
the fact that interviews are recorded. While the participants themselves permit
the recording, they might be extra careful in giving risky statements on the
record and therefore introduce bias in their answers. This threat is minimized
by assuring the recordings themselves will not be published and all results
which will be published are first approved by the corporate communication
department.

\subsubsection{Interpretive Validity}
Interpretive validity concerns errors caused by wrongly interpreting
participants' statements.

% Threats regarding note taking and memoing.
%   - Things could be left out. -> Listening to the audiotape and revising memos.
%   - Things could be perceived incorrectly. -> open-ended follow-up questions (and audiotape listening)
The interviews are processed by field-note taking and memoing. The primary
threat to valid interpretation is imposing one's own meaning, instead of
understanding the viewpoint of the participants and the meanings they attach to
their words. To avoid these interpretation errors, the interviewers used
open-ended follow-up questions which allowed the participant to elaborate on
answers.

\subsubsection{Generalizability}
Generalizability refers to the extent to which one can extend the results to
other settings than those directly studied.

% - Research conducted at a large bank. Results may not be generalizable to orginizations...
%   - ...of different size.
%   - ...of different nature (data driven companies vs. operation driven).
%   - ...prone to less regulations.
%   - ...dealing with less sensitive data. -> GDPR is the same for everyone, and some departments within ING also deal with less sensitive data.
This research is conducted in a single organization -- a large financial institution. Despite being
only one case, we argue that many of the challenges being solved at \ING{} are relatable to
organizations embracing AI into their business. However, results may not seem generalizable to
companies of much smaller size or different nature. A bank may be prone to more regulations than
most companies and is dealing with more sensitive data. Nevertheless, every company has to comply
with privacy regulations like the European GDPR. This suggests that results influenced by more
strict regulations and compliance are just as relatable to other industries. Multiple case studies
at organizations of different scale and nature are required for establishing more general results.

%!TEX root = ../main.tex

\section{Conclusions} \label{sec:conclusions}
% Goal
The goal of this study is to understand the evolution of Machine Learning
development and how state-of-the-art lifecycle models fit the
current needs of the AI industry.
% Method
To that end, we conducted a case study with seventeen Machine Learning
practitioners at the fintech company \ING{}.
% Findings
Our key findings show that traditional Machine Learning lifecycle models are missing essential
steps, such as feasibility study, documentation, model evaluation, and model monitoring. This calls
for more research to aid practitioners in these essential stages.

We also observe that model governance and
technology access are key challenges to the fintech industries leading the AI
revolution. Finally, we have found that existing tools to aid Machine Learning
development do not address the specificities of different projects, and thus,
are seldom adopted by teams.

%Implications
Our research helps practitioners fine-tune their approach to Machine Learning
development to fit fintech use cases. Additionally, it guides educators in
defining learning objectives that meet the current needs in the industry.
%Future Work
Finally, this work paves the way for the next research steps in reducing bottlenecks in the
Machine Learning lifecycle. In particular, it highlights the need for tool support for exploratory
data analysis and data integration techniques, documentation, model governance,
monitoring, and version control.

\ifthenelse{\boolean{anonymous}}{}{%
\section*{Acknowledgments}
The authors would like to thank Irene Wijk, Shiler
Khedri, and Elvan Kula for their willing contributions to this project. The
authors would also like to thank all the
participants of the interviews at \ING{}.%
}

{\footnotesize
\bibliographystyle{spbasic}
\bibliography{bibliography}
}
\end{document}